\documentclass{aa}
\input{psfig}
\def\logg{log {\sl g}}
\def\vs.ini{{\sl v sin~i~}}

\def\kms{${\rm km~s^{-1}}$}

\def\Teff{${\rm T_{eff}~}$}

\begin{document}
\title{New analysis of the two carbon-rich stars
 CS~22948-27 and CS~29497-34: binarity and neutron capture elements
\thanks{Observations collected at the European Southern Observatory
(ESO), ESO Programme 72.A-0244 (PI Petitjean), and
 including data obtained from the ESO/ST-ECF Science Archive
Facility  }}
%
%
%\subtitle{}
%
\author{
B. Barbuy\inst{1}
\and
M. Spite\inst{2}
\and
F. Spite\inst{2}
\and
V. Hill\inst{2}
\and
R. Cayrel\inst{3}
\and
 B. Plez\inst{4}
\and
 P. Petitjean\inst{5}
 }
%
%  \offprint {}
%
\institute{
Universidade de S\~ao Paulo, Rua do Mat\~ao 1226, S\~ao Paulo 05508-900,
 Brazil 
%Email: barbuy@astro.iag.usp.br
\and
Observatoire de Paris-Meudon, GEPI, URM 8633 du CNRS, F-92195 Meudon Cedex, France
%Email: Monique.Spite@obspm.fr, Francois.Spite@obspm.fr, Vanessa.Hill@obspm.fr
\and
Observatoire de Paris, GEPI, 61 Av. de l'Observatoire,
F-75014 Paris, France
%Email: Roger.Cayrel@obspm.fr
\and
GRAAL cc72, Universit\'e de Montpellier II, F-34095 Montpellier Cedex 5, France
%Email: plez@graal.univ-montp2.fr
\and
Institut d'Astrophysique, 98bis Bd. Arago, 75014 Paris, France
%Email: Patrick.Petitjean@iap.fr
}

\abstract{
We have carried out a new determination of abundances in the very
metal-poor CH/CN strong stars CS~29497-34 and CS~22948-27, using
high-resolution spectra obtained with the HARPS spectrograph at the
3.6m telescope of ESO, La Silla, that covers the range
$\lambda\lambda$ 4000 - 6900 {\rm \AA} at a resolution of R = 100,000.
Both stars are found to be long period binaries.  It is confirmed that
the abundance patterns show an enhancement of all the $\alpha$-elements
(like Mg, Ca),  of the proton capture elements (like Na
and Al) and a strong enrichment in "r" and "s" process elements, 
where the s-enrichment is probably
due to a mass transfer episode from a companion in its AGB phase.  The
possible origins of the abundance pattern and especially of the strong
enhancement of both "s" and "r" elements are discussed.

}

\date{}
\authorrunning{B. Barbuy et al.}
\titlerunning{Abundances in CS~22948-27 and  CS~29497-34 }
\maketitle

%--- 1 ---
\section{Introduction.} 

The most metal-poor stars in the Galaxy show an unexpected fraction of
carbon-rich stars, as first pointed out by Beers, Preston \& Shectman
(1992), Norris, Ryan and Beers (1997), Rossi, Beers and Sneden (1999).
More recently, a deeper survey  indicated (Christlieb 2003) 
that about 30\% of the most metal-poor stars are
carbon enhanced.  These stars are often enriched also in nitrogen and
in neutron capture heavy elements.  The enhancement of s-process
and/or r-process elements varies however from star to star.

The enrichment in neutron-capture $s$-elements suggests either an intrinsic
 enrichment in the star itself (post-AGB or helium flash), 
or an extrinsic process, e.g. by mass transfer from a companion
in its Asymptotic Giant Branch (AGB) phase.

An analysis of the elemental abundances of the carbon stars
 BPS~CS~22948-27 and BPS~CS~29497-34  (Barbuy et al.  1997; Hill et al.
2000, Paper I) has shown that these stars are rich in $s$- and $r$-elements.
Only few very metal-poor stars are known with a strong "r" and "s"
enrichment (Zijlstra 2004; Cohen, Christlieb, Qian and Wasserburg 2003,
hereafter CC03).  In our previous work, we were not able to determine the
abundance of lead which is a key element to determine the neutron
exposure since it belongs to the "third peak" of the "s" elements.

Recently new high resolution spectra (R=100,000) of CS~22948-27 and
CS~29497-34 have been obtained at ESO with the high resolution
spectrograph HARPS. In these spectra, several lines appear now less
severely blended and the accuracy of the abundance determination is
much improved.  In this paper we redetermine the abundances of the
elements (and in particular those of Eu and Pb), based on these new
very high resolution spectra.  The opacities and models, computed for
these peculiar chemical compositions, are the same as those used in
the previous paper, whereas revised line lists of CH and CN blue,
and hyperfine structure for the heavy elements
are taken into account.

In order to better
 understand the cause of the overabundance of the heavy elements in
carbon stars, it is important to know whether they are binaries.  In
this case,  an enhancement of the "s" elements for example can be
explained by a transfer of matter from an AGB companion.  In Hill et
al. (2000), from observations in the period 1995-1999, we could not
detect any clear variation of the radial velocity of CS~22948-27 and
CS~29497-34.  However, the radial velocity of CS~22948-27 has been
found recently to be variable by Preston and Sneden (2001).  Using our
additional new data, we have found that the radial velocity of
CS~29497-34 is also variable and we have made a new determination of
the orbital parameters of the two stars.

In Sect.  \ref{Observations} the observations and reduction procedures
are reported and new orbital elements for CS~22948-27 and CS~29497-34
are determined.  In Sect.  \ref{Analysis} the spectrum synthesis
calculations and abundance determinations are described, and the
results are presented.  In Sect.  \ref{Discussion} these results are
discussed and compared to other analyses of heavy element enriched
carbon-stars, in order to try to find possible scenarios 
for the formation of these
peculiar stars.  Finally, concluding remarks are given in Sect.
\ref{Conclusions}.

%-----------------
%--- 2 ---

\section{Observations} \label{Observations}

New high resolution spectra were obtained for CS~22948-27 and
CS~29497-34, in the wavelength region 3800-6900 {\AA}, using the 3.6m
telescope at ESO, equipped with the High Accuracy Planetary Search
Spectrograph (HARPS).  The resolution of these spectra is very high
(R=100,000) and their S/N ratio per pixel (difficult to estimate
because of the crowded lines) is about 50 at 4000 {\AA} and 100 at
6000 {\AA}. Although they have a moderate S/N ratio, their high
resolution is a significant advantage for the identification and
measurement of the lines in these very crowdy spectra.  The HARPS
spectra have been reduced by the corresponding pipeline at the
telescope (division by a flat-field frame, optimal extraction of the
orders, wavelength calibration, and merging of all overlapping
orders).

A spectrum of CS~29497-34 from the ESO archive obtained with UVES at
the VLT-UT2 telescope has also been used for radial velocity
measurements.  This spectrum has a resolution of 45,000, and the
spectral coverage is 3600-4800 {\rm \AA} in the blue and 5880-7400
{\rm \AA} in the red.  This archive UVES spectrum has been reduced
using the UVES context (Ballester et al.  2000) within MIDAS,
and its S/N ratio is close to 200.

The log of the new observations is reported in Table \ref{T-log} (UVES
and HARPS spectra).  In this table we included also our old less
precise observations of these stars from the ESO spectrographs EMMI,
CASPEC and FEROS (see Hill et al. 2000) for the determination of the
orbital parameters.

% TABLE 1
\begin{table}
\caption[1]{Log-book of the observations and Radial Velocities 
measurements} 
\label{T-log}
\begin{flushleft}
\begin{tabular}{lccccc}
\noalign{\smallskip}
\hline
\noalign{\smallskip}
	    & julian d.   & exp.  & $\lambda$ & $\rm Vr_{bary}$\\
R & $-2.4 10^{6}$d & (min) & ($nm$)  & kms$^{-1}$\\
\noalign{\smallskip}
\hline
\noalign{\smallskip}
\multicolumn{3}{c}{\textbf{CS 22948-27}}  \\
\noalign{\smallskip}
{\bf EMMI}   & & & \cr
30,000  & 49976. & 120 & 500-820 &$-63.9\pm 1.5$\cr  
	& 49978. & 105 & 500-820 &$-64.7\pm 1.5$\cr  
	& 50025. &  60 & 500-820 &$-62.3\pm 1.5$\cr 
	& 50026. &  60 & 500-820 &$-62.7\pm 1.5$\cr 
	& 50027. &  55 & 500-820 &$-62.8\pm 1.5$\cr 
{\bf CASP.} & & & \cr
21,000  & 50358. &  90 & 400-500 &$-60.6\pm 3.0$\cr
	& 50359. & 120 & 400-500 &$-67.3\pm 3.0$\cr  
{\bf FEROS}  & & & \cr  
48,000  & 51575. &  30 & 400-850 &$-67.1\pm 1.0$\cr  
	& 51384. &  30 & 400-850 &$-67.4\pm 1.0$\cr  
	& 51394. &  30 & 400-850 &$-68.1\pm 1.0$\cr       
{\bf HARPS}  & & & \cr
100,000 & 53142. &  80 & 380-690 &$-69.9\pm 0.3$\cr
\hline 
\noalign{\smallskip}
\multicolumn{3}{c}{\textbf{CS 29497-34}}  \\
\noalign{\smallskip}
{\bf EMMI}   & & & \cr
30,000  & 49975. & 120 & 500-820 &$-44.6\pm 1.5$\cr 
	& 49975. &  95 & 500-820 &$-44.7\pm 1.5$\cr 
	& 50025. &  90 & 500-820 &$-46.6\pm 1.5$\cr 
	& 50026. &  75 & 500-820 &$-45.7\pm 1.5$\cr 
{\bf CASP.} & & & \cr
21,000  & 50358. & 120 & 400-500 &$-43.3\pm 3.0$\cr 
	& 50359. & 120 & 400-500 &$-42.2\pm 3.0$\cr  
{\bf FEROS}  & & & \cr  
48,000  & 51575. &  30 & 400-850 &$-44.2\pm 1.0$\cr 
	& 51384. &  30 & 400-850 &$-44.5\pm 1.0$\cr  
	& 51394. &  30 & 400-850 &$-44.7\pm 1.0$\cr       
{\bf UVES}   & & & \cr
45,000  & 52472. &  45 & 480-680 &$-51.4\pm 0.5$\cr  
{\bf HARPS}  & & & \cr
100,000 & 52994. & 180 & 380-690 &$-52.7\pm 0.3$\cr
\noalign{\smallskip} 
\hline
\end{tabular}
\end{flushleft} 
\end{table}

\subsection{Radial velocities and orbital elements}

From the new high resolution spectra obtained at UVES and HARPS we
have derived precise measurements of the radial velocity of
CS~22948-27 and CS~29497-34 in 2002, 2003 or 2004.  The barycentric
radial velocities determined from the position of the iron lines are
given for all the spectra in Table \ref{T-log}. 
 It is difficult to
find unblended lines in these cool stars and the formal standard
deviation of the mean radial velocity is 0.27 and 0.28 km s$^{-1}$ 
for the HARPS and UVES spectra respectively.
% Table 2
\begin{table}
\caption[1]{Provisional orbital elements for CS~22948-27 and CS~29497-34}
\label{T-orbit}
\begin{center}
\begin{tabular}{l@{ }c@{ }c@{ }c@{  }c@{}c}
\noalign{\smallskip}
\hline
\noalign{\smallskip}
Element & CS~22948-27 & CS~29497-34 \\
\noalign{\smallskip}
\hline
JDo            & 2448110& 2449800 \\
$\gamma$ (\kms)&  -67.5 &  -47.5  \\
K1 (\kms)      &   4.4  &  5.2    \\
e              &  0.02  &  0.02   \\
$\omega$ (rad) & 0.011  &  0.07   \\
P (days)       & 426.5  &  4130   \\
\noalign{\smallskip} 
\hline 
\end{tabular}
\end{center}
\end{table}

\subsubsection{CS~22948-27}
\begin{figure}[ht]
\psfig{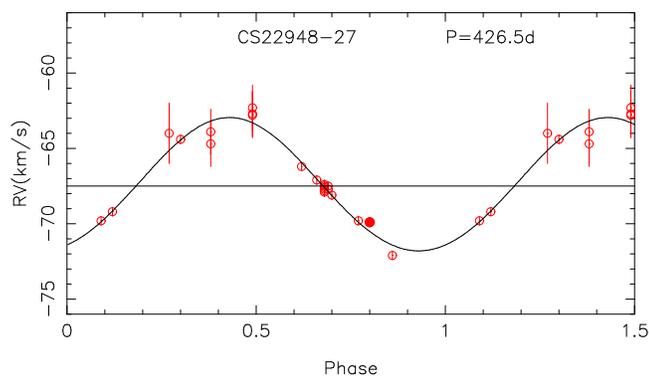}
\caption {Orbital variation of the radial velocity of 
CS~22948.27 computed from the  orbital 
elements in Table \ref{T-orbit} (the new observations are marked 
with filled circles). The period is close to 426.5d.}
\label{F-vr22948}
\end{figure}

Recently Preston and Sneden (2001) have monitored the radial velocity
of CS~22948-27 at Las Campanas observatory and found it variable with
a period of 501 days, but the data were sparse and the period
uncertain.  Our measurement of the radial velocity in 2004 cannot be
reconciled with the curve of the orbital variations of the radial
velocity given by Preston and Sneden (at an orbital phase $\phi$=0.96
we find RV=-69.9$\pm0.3$ \kms, whereas
 from the published curve we would expect
-65.0 \kms).  However, in the determination of the orbital parameters
Preston and Sneden (2001) had rejected two Las~Campanas observations
(JD~50997 RV=-69.8 and JD~51034 RV=-72.1 \kms).  We supplemented our own
ESO observations of CS22948-27 with {\sl all} the Las Campanas
Observatory data.  We have replaced our two imprecise CASPEC
measurements (see Table \ref{T-log}) by their mean value.  The orbital
elements of CS~22948-27 (Table \ref{T-orbit}) were derived with the
code "velocity" (Wichmann et al. 2003).  The new period, achieving a
satisfactory agreement of the individual measurements with the
computed curve, has been found to be close to 426d.  The new velocity
curve generated with the orbital elements listed in Table
\ref{T-orbit} is displayed in Figure \ref{F-vr22948}.

%Fig. 1

\subsubsection{CS~29497-34}
\begin{figure}[ht]
\psfig{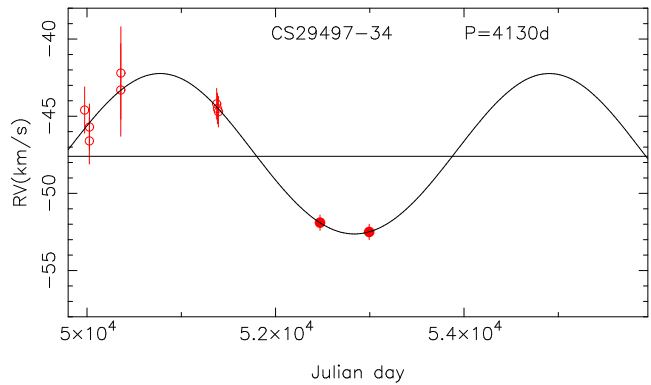}
\caption {Radial velocities of CS~29497-34 measured as a
function of time.
The abscissa is the Julian date -- 2 400 000d.  The curve was
generated by the orbital elements listed in Table \ref{T-orbit} (the
new observations are marked with filled circles).  The period is close
to 4130d.}
\label{F-vr29497}
\end{figure}

In Fig. \ref{F-vr29497} we present the sequence of radial velocity
measurements from 1995 to 2004 for CS 29497-34.
  From the last two values it appears
now clearly that the radial velocity of CS~29497-34 is also variable.
The orbital elements derived with the code "velocity" (Wichmann et
al. 2003) are given in Table~\ref{T-orbit}.  The period has been found
to be close to 4130 days.  However the uncertainty is very large and
more observations are needed to obtain a better precision in the
determination of the orbital elements.

Finally, it is now established that the carbon-rich stars CS~22948-27 
and CS~29497-34 are both binaries, both with long periods.

%Fig. 2

%--------------------------------------------------------------------
%--- 3 ---

% Table 3
\begin{table}
\caption[1]{Stellar parameters effective temperatures, surface
gravities, metallicities, and microturbulence velocities, as
derived in Hill et al.  (2000).}
\label{T-parameters}
\begin{flushleft}
\begin{tabular}{l@{ }c@{ }c@{ }c@{  }c@{}c}
\noalign{\smallskip}
\hline
\noalign{\smallskip}
& \Teff   &\logg & ~~${\rm [Fe_{I}/H]}$~~ & ~~${\rm [Fe_{II}/H]}$~~ &  $v_t$  \\
&     (K) &      &                        &              & ~~km/s~~\\
\noalign{\smallskip}
\hline
\noalign{\smallskip}
CS~22948-27 & 4800 & 1.8 &  $-$2.47 &  $-$2.46 & 1.5 \\
CS~29497-34 & 4800 & 1.8 &  $-$2.90 &  $-$2.91 & 1.5 \\
\noalign{\smallskip} \hline \end{tabular}
\end{flushleft}
\end{table}

\section{Analysis} \label{Analysis}

\subsection{Model Atmospheres}

We used model atmospheres provided by the OSMARCS code, where LTE
equilibrium, homogeneity and conservation of the total flux are
assumed.  It has been developed by Gustafsson et al.  (1975) and later
improved by Plez et al.  (1992), Edvardssson et al.  (1993), Asplund
et al.  (1997) and Gustafsson et al.  (2003).  The models take into
account the enhancement of the carbon and nitrogen abundances in the
atmosphere.  The atmospheric parameters (effective temperature T$_{\rm
eff}$, surface gravity log g, metallicity [Fe/H] and microturbulence
velocity v$_{\rm t}$), adopted from Paper I, are given for
convenience in Table \ref{T-parameters}.

% Table 4
\begin{table}[ht]
\caption[1]{Abundances of the elements in CS~22948-27 and 
CS~29497-34.}
\label{T-metalabund}
\begin{center}
\begin{tabular}{lccccc}
\hline
\noalign{\smallskip}
  elem & log $\epsilon_{*}$ & [M/H] & [M/Fe] & $\sigma$ & N\\
\noalign{\smallskip}
\hline
\noalign{\smallskip}
\multicolumn{5}{c}{\textbf{CS 22948-27 [Fe/H]=-2.47} } \\
\noalign{\smallskip}
C  &  8.49   & -0.03     & +2.43  &  & synt\\
N     & 7.21     & -0.71 & +1.75 & & synt\\
 Na I* & 4.43 & -1.90 & +0.57 & 0.17 & 2\\
 Mg I & 5.42 & -2.16 & +0.31 & 0.28 & 4\\
 Al I* & 3.85 & -2.62 & -0.15 &      & 1\\
 Ca I & 4.43 & -1.93 & +0.54 & 0.15 & 9\\
 Ti I & 2.89 & -2.13 & +0.34 & 0.28 &10\\
 TiII & 3.05 & -1.97 & +0.50 & 0.20 &20\\
 Cr I & 3.05 & -2.62 & -0.15 & 0.10 & 3\\
 Ni I & 3.77 & -2.48 & -0.01 &      & 1\\
 Zn I & 2.49 & -2.11 & +0.36 &      & 1\\
 Sr II& 1.35 & -1.57 & +0.90 &      &synt\\ 
 Y  II& 0.87 & -1.47 & +1.00 &      &synt\\  
 Ba II& 1.92 & -0.21 & +2.26 &      &synt\\ 
 La II& 1.02 & -0.15 & +2.32 &      &synt\\  
 Ce II& 1.31 & -0.27 & +2.20 &      &synt\\  
 Pr II& 0.08 & -0.63 & +1.87 &      &synt\\  
 Nd II& 1.25 & -0.25 & +2.22 &      &synt\\  
 Eu II&-0.08 & -0.59 & +1.88 &      &synt\\  
 Dy II& 0.27 & -0.87 & +1.60 &      &synt\\ 
 Pb I & 2.20 & +0.25 & +2.72 &      &synt\\  
\noalign{\smallskip} 
\hline  
\noalign{\smallskip}
\multicolumn{5}{c}{\textbf{CS 29497-34 [Fe/H]=-2.90} } \\
\noalign{\smallskip}
C  &  8.24   & -0.28     & +2.63  &  & synt\\
N     & 7.39   &  -0.53   &  +2.38     &     & synt\\
Na I* & 4.61 & -1.72 & +1.18 & 0.19 & 2\\
Mg I & 5.40 & -2.18 & +0.72 & 0.34 & 4\\
Al I* & 3.56 & -2.91 & -0.01 &      & 1\\
Ca I & 3.91 & -2.45 & +0.45 & 0.18 & 9\\
Ti I & 2.41 & -2.61 & +0.29 & 0.28 &11\\
TiII & 2.56 & -2.46 & +0.44 & 0.27 &15\\
Cr I & 2.54 & -3.13 & -0.23 & 0.12 & 3\\
Ni I & 3.36 & -2.89 & +0.01 &      & 1\\
Zn I & 2.15 & -2.44 & +0.46 &      & 1\\
Sr II& 1.02 & -1.90 & +1.00 &      &synt\\ 
Y  II& 0.54 & -1.80 & +1.10 &      &synt\\  
Ba II& 1.26 & -0.87 & +2.03 &      &synt\\ 
La II& 0.39 & -0.78 & +2.12 &      &synt\\  
Ce II& 0.63 & -0.95 & +1.95 &      &synt\\  
Pr II&-0.37 & -1.08 & +1.82 &      &synt\\  
Nd II& 0.69 & -0.81 & +2.09 &      &synt\\  
Eu II&-0.59 & -1.10 & +1.80 &      &synt\\  
Dy II&-0.26 & -1.40 & +1.50 &      &synt\\ 
Pb I & 2.00 & +0.05 & +2.95 &      &synt\\  
\noalign{\smallskip} \hline 
\multicolumn{6}{l}{*To take into account the NLTE effects the abundances of}\\
\multicolumn{6}{l}{~ Na given in the table have to be decreased by 0.5dex and }\\ 
\multicolumn{6}{l}{~ the abundances of Al have to be increased by 0.65dex}\\
\end{tabular}
\end{center}
\end{table}

\subsection{Carbon and Nitrogen abundance}
A new determination of the abundance of carbon and nitrogen
has been performed, using the CH lines based on line list
by Luque \& Crosley (1999), and CN red lines as previously.\\
We assume abundances $\epsilon$ (C) = 8.52 and $\epsilon$ (N) = 7.92
(Grevesse \& Sauval 1998). The abundances obtained are
 [C/Fe] = +2.43 and [N/Fe] = +1.75 for CS~22948-27
and   [C/Fe] = +2.63 and [N/Fe] = +2.38 for CS~29497-34.
(Table \ref {T-metalabund}).

\subsection{Abundances of metals from Na to Zn}
In Barbuy et al.  (1997) and Hill et al.  (2000), due to
crowding of lines, we could only
determine the abundances of C, N, Fe, Na and Mg and 
 heavy elements.  In the present paper the new higher
resolution spectra of CS~22948-27 and CS~29497-34
allowed  to measure 
equivalent widths of metallic lines from Na to Zn.  This abundance
analysis was performed using Turbospectrum (Alvarez \& Plez 1998)
and the results are given in Table \ref{T-metalabund}. 
Solar abundances for the elemental species were adopted from Grevesse
\& Sauval (1998).

The sodium and the aluminum abundances are computed from the resonance
doublets: Na D lines at 5890 and 5896 {\AA} and the Al line at 3961.5
{\AA}
(the other line of the aluminum doublet is too severely blended to be
used).  All these lines are sensitive to NLTE effects (Cayrel et
al. 2004; Baum\"uller et al.  1998; Baum\"uller and Gehren 1997;
Norris et al. 2001).  As a consequence the abundance of sodium given
in Table \ref{T-metalabund} should be decreased by 0.5 dex and that
of aluminum should be increased by about 0.65dex
(in Fig. 3 the corrected values are used).

%Fig. 3
\begin{figure}[]
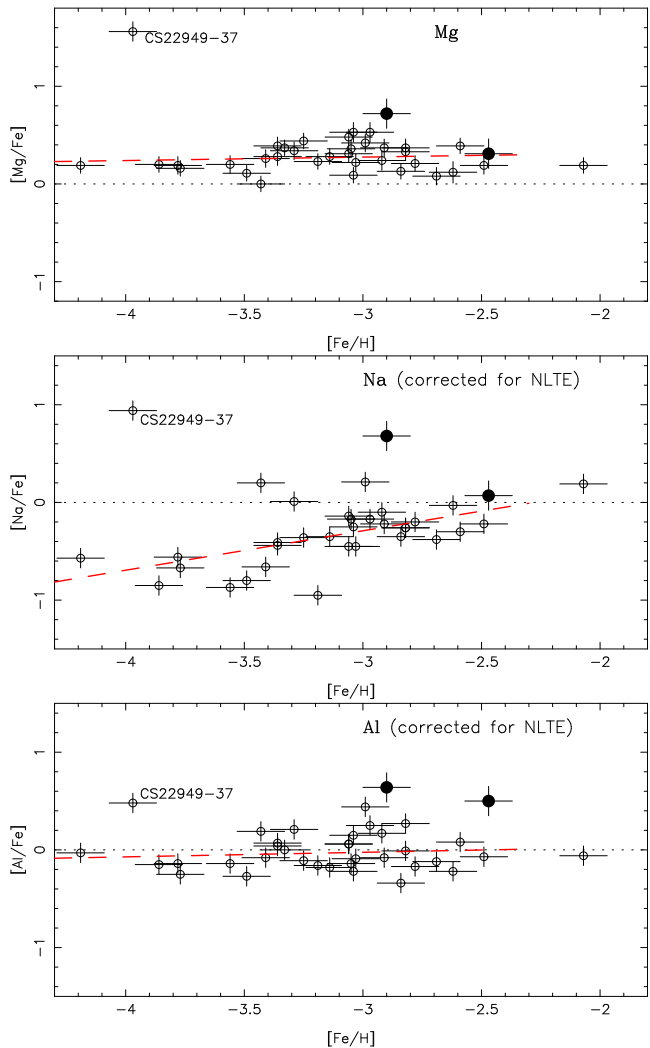
 
\psfig{file=abgfe-mgchcn.ps,angle=0,width=8.5 cm}
\psfig{file=abgfe-na2chcn.ps,angle=0,width=8.5 cm}
\psfig{file=abgfe-al2chcn.ps,angle=0,width=8.5 cm}
\caption {Abundances of Mg, Na and Al in CS~22948-27 ([Fe/H]=--2.47) and 
CS~29497-34 ([Fe/H]=--2.90) (filled circles) compared to the abundances
of these elements measured in extremely metal-poor stars
(open circles). CS~22949-37 (Depagne et al., 2002) is also a carbon 
rich star but its very peculiar pattern is explained by the ejecta of 
a zero-metal SNII with large fallback.}
\label{F-abmet}
\end{figure}

We have compared the abundances of metals in CS~22948-27 and 
CS~29497-34 to the abundances found in 
Cayrel et al. (2004) for extremely metal poor stars
(Fig.~\ref{F-abmet}).
There is  rather good agreement for Ca, Ti, Cr, Ni and Zn,
whereas  CS~29497-34 is Mg-rich, Na-rich and Al-rich.
In  CS~22948-27 the abundance of Mg and Na are pratically as expected
for a normal star with [Fe/H]=--2.5 (i. e. showing the classical enhancement of the 
$\alpha$-elements), whereas the abundance of Al is enhanced.

Fig. 3 shows [Mg/Fe], {Na/Fe] and [Al/Fe] vs. [Fe/H]
 in CS~22948-27 and CS~29497-34,  compared with the abundances
of these elements measured in extremely metal-poor stars.
In this Fig.,  CS~22949-37 (Depagne et al. 2002) is also
plotted illustrating its peculiar characteristics: it is carbon 
rich, and its pattern can be explained as
typical  ejecta of a  massive zero-metal SNII with large fallback,
 according to Woosley \& Weaver (1995), and a Z = 10$^{-4}$
according to (Umeda \& Nomoto 2003).

\subsection{Heavy element abundances}
The  heavy element abundances  have been  determined 
by spectrum synthesis calculations. The codes 
Turbospectrum described by Alvarez and Plez (1998), and that 
described by Barbuy et al. (2003) were both used. 

Absorption lines of the following molecules were taken into account in the
calculations: MgH (A$^2 \Pi$-X$^2 \Sigma$), C$_2$ (A$^3 \Pi$-X$^3 \Pi$), CN
blue (B$^2 \Sigma$-X$^2 \Sigma$), CH (A$^2 \Delta$-X$^2 \Pi$), CH (B$^2
\Delta$-X$^2 \Pi$), and CN red (A$^2 \Pi$-X$^2 \Sigma$).

For lines of the heavy elements BaII, LaII and EuII, a hyperfine
structure was taken into account, based on the hyperfine constants
by Lawler et al. (2001a) for EuII, Lawler et al. (2001b) for LaII
and Biehl (1976) and McWilliam (1998) for BaII.
For the PbI 4057.807 line,
the hyperfine structure given in van Eck et al. (2003)
was adopted.

For most of the heavy elements our new determinations
 (Table~\ref{T-metalabund}) confirm the
values found by Paper I.  
For the determination of the Pr abundance we could add the blue
lines at 3964.3, 3964.8 and 3965.3 {\AA} and we found an abundance 
slightly higher than that of Paper I.
Moreover we revised the
abundance of europium and we could determine, for the first time, the
abundance of Pb in these stars.

\subsubsection{Eu}
The abundance of Eu in these cool carbon-rich stars is best 
determined from the red lines at 6437.64 and 6645.13 {\AA} 
but an unidentified line  appears more clearly in the present
data on the blue side of the 6645.13 {\AA} line, with the result 
that, relative to Paper I, [Eu/Fe] is decreased from
+2.10 to +1.88 in CS 22948-27 and from +2.25 to +1.8
in CS 29497-34 (Fig.~\ref {F-Eufit}). 

We tried to use the blue europium lines at 4129 and 4205 {\AA} to
estimate its isotopic ratio.  In s-process
element-enhanced metal-poor stars, this ratio has been always
found to be close to the solar value (see Sneden et al.  2002; Aoki
et al. 2003, hereafter AR03), 
but it would be interesting to check this ratio in r
{\sl and} s-process enhanced stars like CS~22948-27 and CS~29497-34.
However the blue EuII lines in these stars are severely
blended with strong CN features, and do not allow the derivation
of  isotopic ratios.

%Fig.4 
\begin{figure}[ht]
\psfig{file=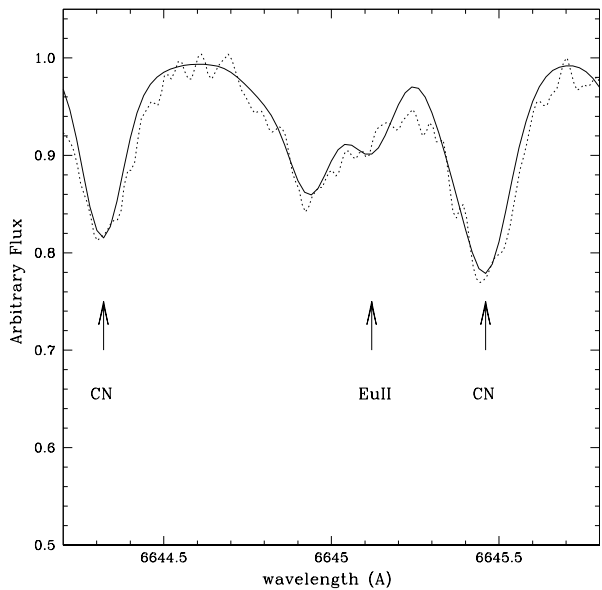,angle=0,width=8.5 cm}
\caption {CS 29497-34:
EuII 6645 line computed with log $\epsilon(Eu)=-0.60$ i.e. [Eu/Fe]=+1.8.
}
\label{F-Eufit}
\end{figure}

\subsubsection{Pb}
We were able to determine the abundance of Pb, thanks to the
high resolving power of the present observations.
Fig.~\ref{F-Pbfit} shows the fit to the PbI 4057.8 {\AA} line, in
both stars.
The overabundance is very large: the best fit  in CS~22948-27 is obtained for 
log $\epsilon (Pb)= +2.2 $  
(i.e. [Pb/Fe] = +2.72) and 
in CS~29497-34  for log $\epsilon (Pb)=+2.00    $  (i.e. [Pb/Fe] = 
+2.95).

%Fig.5
\begin{figure}[ht]
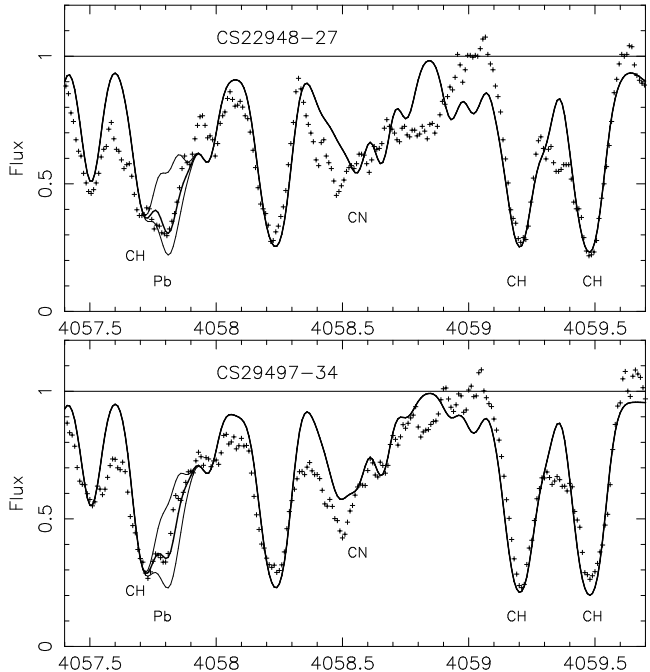

\psfig{file=pb22948-27.ps,angle=0,width=8.5 cm}
\psfig{file=pb29497-34.ps,angle=0,width=8.5 cm}
\caption {Pb~I line computed with log $\epsilon (Pb) =$ 1.5, 2.5
and the adopted value (log $\epsilon (Pb) =$ 2.2 for CS~22948-27, 
and log$\epsilon_{Pb} =$ 2.0 for CS~29497-34.
These values correspond to [Pb/Fe]$_{22948-27} = +2.7$  and 
[Pb/Fe]$_{29497-34} = +2.95$
}
\label{F-Pbfit}
\end{figure}

       \subsubsection{General pattern of the heavy elements} 
       In Fig.~\ref{F-HeavyPat} the patterns of the heavy elements 
       relative to Ba in CS~22948-27 and CS~29497-34 are compared to the r- 
       and s-process solar abundances from Burris et al. (2000). The 
       abundances of the lighter 
       elements Sr and Y seem to be  discrepant: the enhancements in Sr
      and Y are much lower than the enhancements in Ba (Table~\ref{T-metalabund}) 
      and consequently the ratios Y/Ba and Sr/Ba appear as particularly low. 
      On the contrary, the enhancement of Pb is larger than the enhancement of Ba.
      For the elements heavier than Ba (excluding Pb), the pattern
     of the elements is close to the solar "s only curve", 
     but generally a little higher. Europium in particular is
     very rich  in CS~22948-27 and CS~29497-34, indicating that the observed
   pattern implies a contribution from both s- and r-processes.
 The numerical values of the observed ratios [Ba/Eu]
 (see Table~\ref{T-abCHCN}), intermediate between the values for 
 r-only- and s-only process in the solar system, lead to the same 
conclusion. Owing to the deviations of Sr, Y and Pb, the general pattern
      is clearly non-solar. These peculiarities are also observed in similar
      stars (AR02; CC03).  One  of the causes of the 
non-solar pattern is the metal deficiency of these stars, implying
 a higher ratio of neutrons relative to seed nuclei
      (higher neutron flux). 
       Quantitatively, an indicator of the neutron flux [hs/ls] 
       (see Table~\ref{T-abCHCN}) is defined 
       as the ratio of the heavy-  to the light-'s' elements, 
       relative to the solar ratio. It is computed as in 
       Norris et al. (1997) as 
       $<$[Ba/Fe],[Ce/Fe],[Nd/Fe]$>$/$<$[Sr/Fe],[Y/Fe],[Zr/Fe]$>$; 
       note that Zr has been omitted in our calculation of this quantity. 
       The theoretical models predict 
       an increase of the [hs/ls] ratio with decreasing metallicity [Fe/H], 
       and Busso et al. (2001) discuss the limitations of this correlation. 
       %%% 

%Fig.6
\begin{figure}[ht]
\psfig{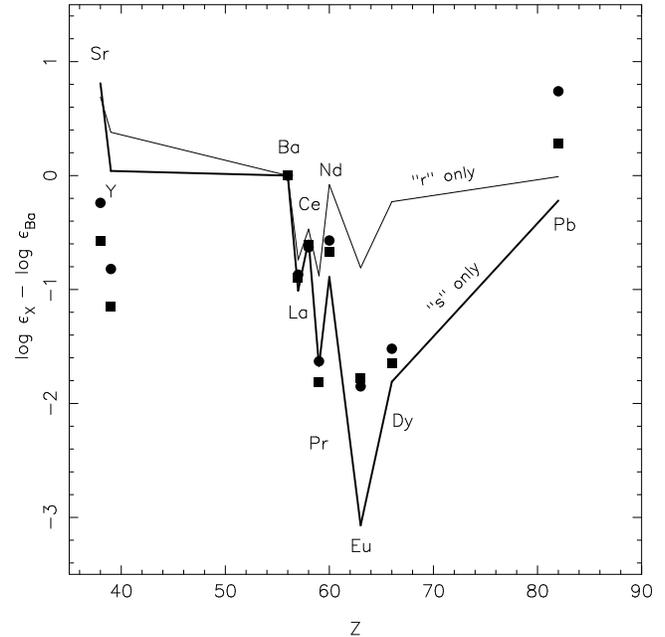}
\caption {Abundances relative to Ba [X/Ba] in CS~22948-27 (filled
square) and CS~29497-34 (filled circle) is compared to the r- and
s-process solar abundances (thin and thick lines) from Burris et al.
(2000). Note that in this figure we have plotted only the pattern of 
the elements really measured in the stars.}
\label{F-HeavyPat}
\end{figure}

% Table 5
\begin{table*}
\caption{Metallicities and abundances of Ba, Pb and Eu, and the
[hs/ls] measure of second peak to first peak s-elements, in selected
very metal-poor stars ([Fe/H] $<$ -2.2), showing enhancements of both
s- and r-process elements, and carbon-rich.  HE~0024-2523 is a
metal-poor s-rich star given here for comparison.
The ratios take into account a solar Fe abundance of 7.50, and a 
renormalization of results in the literature was applied.
  The references are
given in the last column. }
\label{T-abCHCN}
\begin{flushleft}
\begin{tabular}{lllllllllll}
\hline
   Star       & [Fe/H] &[Ba/Fe] &[La/Fe]&[Ce/Fe]&[Eu/Fe]  &[Pb/Fe]& [hs/ls] & ref.\\
 \hline
 \multicolumn{4}{l}{\textbf{"s-" and "r-rich" stars} }\\
 CS~22948-27  & -2.47  &  2.26  & 2.32  &  2.20 &   1.88   & 2.72 &  0.97   & 1,2\\
 CS~29497-34  & -2.90  &  2.03  & 2.12  &  1.95 &   1.80   & 2.95 &  1.28   & 1,2\\
 \\
 HE~2148-1247 & -2.30  &  2.36  & 2.38  &  2.28 &   1.98   & 3.12 &  1.76   & 3  \\
 \\
 CS~29526-110 & -2.38  &  2.20  & 1.74  &  2.06 &   1.77   & 3.43 &    -    & 4  \\
 CS~22898-027 & -2.25  &  2.33  & 2.19  &  2.19 &   1.93   & 2.96 &  1.37   & 4  \\
 CS~31062-012 & -2.55  &  2.07  & 2.07  &  2.17 &   1.66   & 2.50 &  1.51   & 4  \\
 CS~31062-50  & -2.32  &  2.40  & 2.49  &  2.15 &   1.88   & 2.97 &    -    & 4  \\
 HD~196944    & -2.25  &  1.20  & 0.97  &  1.07 &   0.22   & 2.01 &  0.29   & 4  \\
 CS~30301-015 & -2.64  &  1.53  & 0.88  &  1.20 &   0.22   & 1.78 &  0.99   & 4  \\
 LP~625-44    & -2.71  &  2.84  & 2.52  &  2.33 &   2.00   & 2.66 &  1.01   & 5  \\
 LP~625-44    & -2.72  &  2.81  & 2.40  &  2.22 &   1.72   & 2.60 &  1.30   & 6  \\
 LP~706-7     & -2.74  &  2.11  & 1.87  &  1.92 &   1.45   & 2.40 &  1.76   & 5  \\
 \\
 \hline
 \multicolumn{4}{l}{\textbf{"s-rich star} }\\
 HE~0024-2523 & -2.70  &  1.46  & 1.80  &   -   &$<$1.10   & 3.30 &   -     & 7  \\
\hline
 \end{tabular}
{\footnotesize \\
References: 1 present work; 2 Paper I;  3 CC03;
4 Aoki et al. (2002a); 5 AR01; 6 AR02
7 Lucatello et al. (2003) }
\end{flushleft}
\end{table*}

       \section{Discussion of the abundances}\label{Discussion} 

       It is interesting to compare our stars to other stars found in the 
       literature for being  rich in both s- (Pb-rich) and  
       r-process elements, in particular HE~2148-1247 analyzed by 
       CC03, and the stars homogeneously analyzed by 
       Aoki et al. (2001, 2002a,b, hereafter AR01 and AR02). From
       the sample of metal-poor, carbon-rich and Pb-rich stars 
       of AR01 and AR02, we have disconsidered two stars : 
       one because of its relatively high metallicity (we kept the stars with 
       [Fe/H] $<$ -2.2) and the other because 
       only an upper limit of the Pb abundance had been determined.  All 
       these stars are both s-rich and r-rich : we will designate these stars 
       by "r+s stars". The first 4 stars of AR02 in Table 5 are the most
       r-rich, according to their [Ba/Eu] ratio (see however AR03), 
       two of them have variable radial velocities.

       In Table \ref{T-abCHCN} the abundances of Ba, La, Ce, Eu and Pb in 
       CS~22948-27 and CS~29497-34 are listed, along with those of 
       HE~2148-1247 (CC03) and  the AR01 and AR02 ones. 
       One of the common points of the stars in Table \ref {T-abCHCN}, 
       enhanced in r- and s-process elements, is a large abundance of Pb. 
       The metal-poor Pb-rich stars are not many: a recent review is given in 
       Sivarani et al.  (2004), where 25 such stars are reported to be known 
       (but not all of them are "r-rich"). 
       In Table \ref {T-abCHCN} we also added, for comparison purposes, a very 
       metal-poor star dominated by the s-process and not r-rich: HE~024-2523 
       (Lucatello et al. 2003). 

       The two stars analyzed here are very similar to  HE~2148-1247 
       (CC03).  These three stars are long period 
       binaries (and presumably also the third one), they are metal-poor, 
       enhanced        in carbon, 
       nitrogen, $\alpha$ elements, r-process and s-process 
       elements (especially in Pb). They have measured abundances for a number
       of elements and form an homogeneous group 

       %  Fig.7  (ex Fig 5 Beatriz) 
       %%%% 
\begin{figure}[ht]
\psfig{file=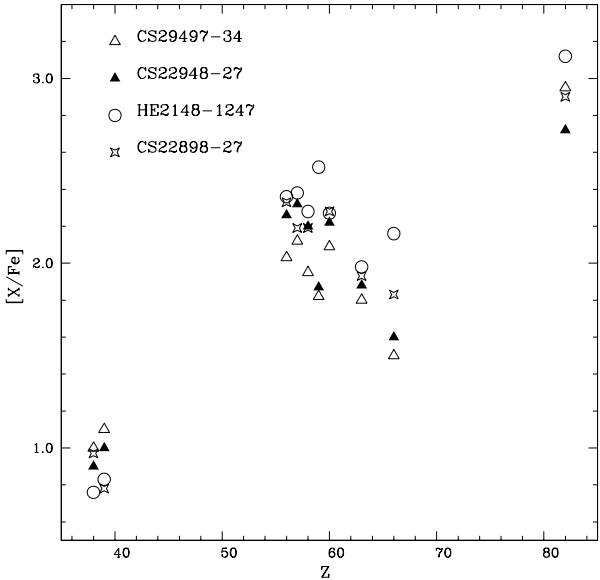,angle=0,width=8.5 cm}
\caption {
Abundances of the heavy elements in CS~22948-27 (filled triangles) 
and CS~29497-34 (open triangles), compared to HE~148-1247 (open 
circles) and one 
of the  AR01, AR02 stars CS~22898-27 (open stars). All the stars here have a
very similar pattern.
}
\label{F-pat-stars}
\end{figure}

       In  Fig. \ref {F-pat-stars} 
       the abundance patterns of CS~22948-27 and CS~29497-34 are compared to 
       those of HE~148-1247 (CC03) and  one star from AR02: CS 22898-27 
       (the eight AR01, AR02 stars in Table \ref{T-abCHCN} 
       are essentially similar and we chose to plot only one star for 
       the lisibility of the figure).

       In these stars the s-process elements have probably been transferred 
       from an AGB companion.  In very metal-poor AGBs, the s-process operates 
       with a high ratio of neutrons to seed nuclei that should favor a high 
       abundance of Pb.  In fact, according to Gallino et al.  (1998), a 
       large production of the double magic nucleus $^{208}$Pb is due to AGB 
       stars of low metallicity, which explains the presence of Pb in 
       low-metallicity stars, as well as the fraction of this isotope in the 
       Sun.  This is confirmed by Travaglio et al.  (2001), together with the 
       fact that, on the other hand, low mass AGB stars (M $\leq$ 4 
       M$_{\odot}$) play the dominant role in the production of s-elements in 
       the Galaxy. Calculations made by Goriely \& Mowlawi (2000) and Goriely
       \& Siess (2001) predict as well large enhancements of third peak
       s-elements relative to first and second peaks.
       The enhancement of Pb in metal-poor stars has been later 
       confirmed by observations (Van Eck 2001). 

       \subsection{Neutron flux and Pb abundance} 

       %Fig. 8 
       %%% 
\begin{figure}[ht]
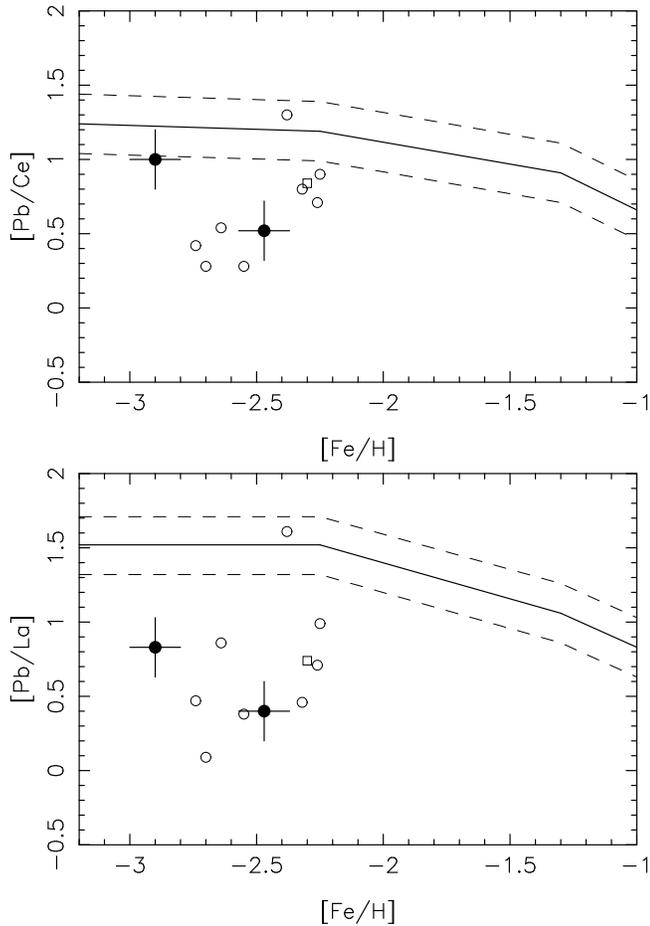

\psfig{file=gori-pb-ce.ps,angle=0,width=8.5 cm}
\psfig{file=gori-pb-la.ps,angle=0,width=8.5 cm}
\caption {[Pb/Ce] and [Pb/La] versus [Fe/H] for CS~22948-27 and
CS~29497-34 (filled circles), HE~148-1247 (open square), 
and the AR01, AR02, AR03 (open circles)
 selected in Table \ref{T-abCHCN}.  Predictions from
the standard PMP model from Goriely and Mowlavi (2000) and Goriely and
Siess (2001) are represented by the solid line.  The dashed line
provide a rough estimate of the uncertainties of these predictions
(see van Eck et al.  2003).  The model predicts generally a too high
ratio of [Pb/Ce] and [Pb/La].  }
\label{F-Goriely}
\end{figure}

       Fig. \ref {F-Goriely}  compares the observed Pb/Ce ratios 
       of our stars, together with the other "r-s-rich" stars in 
       Table~\ref{T-abCHCN}, with
       model predictions of the ejecta of metal-poor AGBs 
       (Van Eck et al. 2003; Goriely and Mowlavi 2000; Goriely and Siess 
       2001). 
       These models predict too high values of [Pb/Ce] 
       and [Pb/La] (see Fig. \ref{F-Goriely}) . The same conclusion was 
       reached by van Eck et al. (2003) 
       for classical "s-rich" metal-poor stars. 

       %  Fig. 9 
       %%% 
% (ex Fig 6 Beatriz) 
\begin{figure}[ht]
\psfig{file=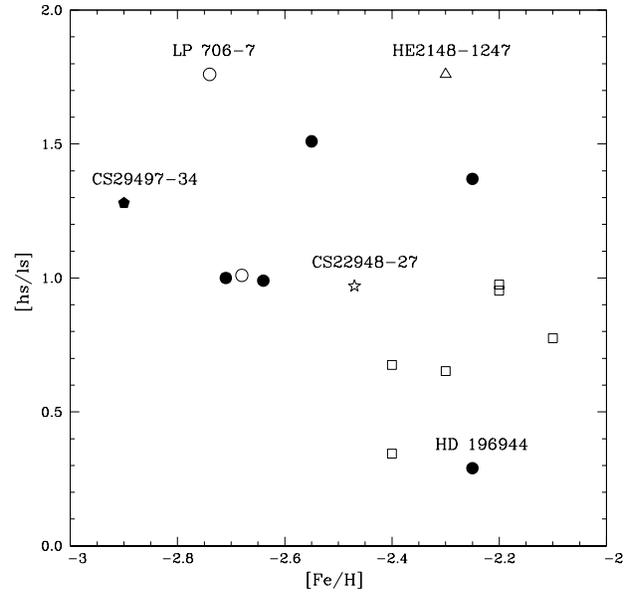,angle=0,width=8.5 cm}
\caption {
[hs/ls] vs. [Fe/H] for Pb-rich stars. Symbols: filled 
pentagone:CS~29497-34, open star: CS~22948-27, open triangle 
HE~2148-1247 (CC03),
open circles: AR01, filled circles: AR02 
al. (2002),  open squares:
van Eck et al. (2003).
}
\label{F-hs-ls}
\end{figure}

       %  Fig. 10 
       % (ex Fig 7 Beatriz) 
\begin{figure}[ht]
\psfig{file=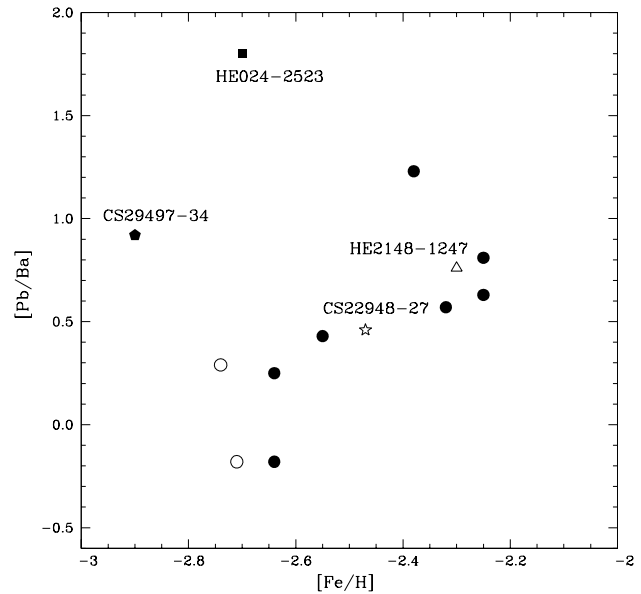,angle=0,width=8.5 cm}
\caption {
[Pb/Ba] vs. [Fe/H] for Pb-rich stars. Symbols: filled square: 
HE~024-2523 (Lucatello et al.2003), and others are the same as in 
Fig. \ref {F-hs-ls}. 
}
\label{F-pb-ba}
\end{figure}

       %  Fig. 11 
       % (ex Fig 8 Beatriz) 
\begin{figure}[ht]
\psfig{file=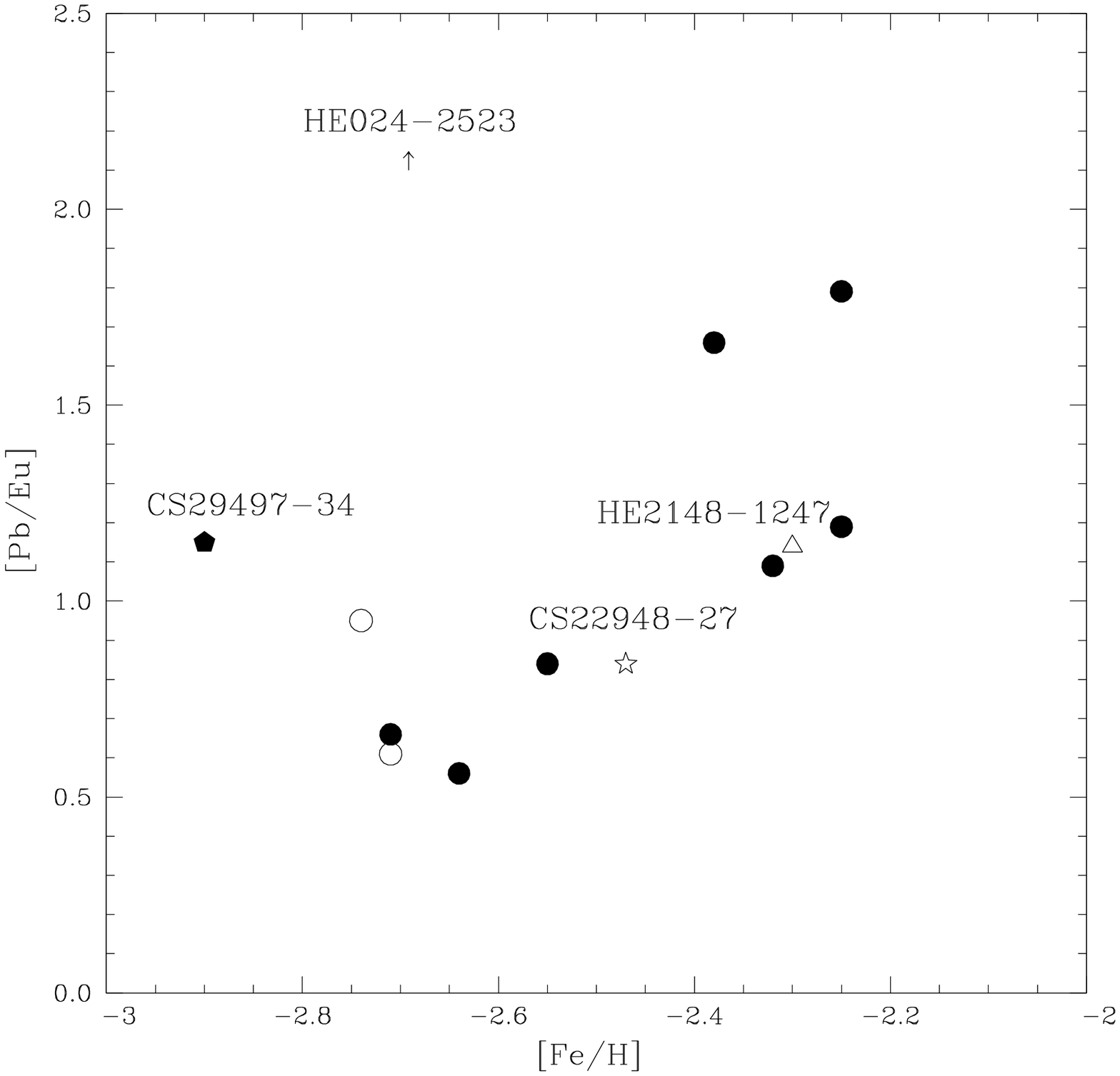,angle=0,width=8.5 cm}
\caption {
[Pb/Eu] vs. [Fe/H] for Pb-rich stars. Symbols: same as in 
Fig. \ref {F-hs-ls}.}
\label{F-pb-eu}
\end{figure}

       In Fig. \ref{F-hs-ls}, [hs/ls] is plotted vs.  [Fe/H] for the Pb and 
       r-rich 
       stars reported in Table \ref{T-abCHCN}.  No clear correlation is 
       observed.  For most of the stars [hs/ls] seems to increase with decreasing 
       metallicity but there are exceptions: HD196944 has a very low [hs/ls] 
       for its metallicity and on the contrary [hs/ls] is very high in 
       LP~706-7. 

       Figs. \ref {F-pb-ba} and  \ref {F-pb-eu} show [Pb/Ba] vs. [Fe/H] 
       and [Pb/Eu] vs. [Fe/H] respectively, for the sample of Table 
       \ref{T-abCHCN}. 
       Whereas there might be a correlation of lead-to-barium 
       (representing a third peak to a second peak s element ratio) with 
       metallicity, the same is not true for the lead-to-europium 
       (a third peak s-element relative to an r-element) ratio. 
       The absence (or the low degree) of correlation indicates that the 
       amount of Pb produced varies from one site to another, and 
        from star-to-star, as noted by CC03 and explained by 
        models (e. g. Busso et al. 1999). 

       A close look at the ratios [Pb/Ce] versus [hs/ls] shows the lack of a 
       clear correlation : for example, compared to LP706-7, the other star 
       (LP 625-44) of AR01 has a Pb/Ce ratio {\it larger} by a 
       factor of ten, but a ratio hs/ls {\it smaller} by a factor of 5, 
       showing for 
       these two stars an anticorrelation. The neutron flux, derived from 
       the comparison of elements of the first and second peak, has  
       no clear correlation with the abundance of Pb (a third-peak 
       element).
       More generally, no clear correlations were found between the Pb 
       abundance and other basic parameters, owing to the sensitivity of 
       this abundance to a number of critical parameters (for example the 
       concentration and the profile of the $^{13}$C pocket,
       e.g. Busso et al. 1999). 

       %section 4.2. 
       \subsection{The heavy elements in metal-poor stars} 

       In normal stars,
       the evolution of  heavy element abundances vs. iron 
        (Burris, Pilachowski, Armandroff et al.  2000) shows 
       that the onset of the r-process (attributed to low mass SN II) is 
       observed at 
       about [Fe/H]~=~-2.9 and that of the s-process (attributed to massive AGBs) 
       at [Fe/H]~=~-2.75, near the metallicities of the two stars analysed 
       here (and the one by CC03). 
        The ratios X/Fe (where X means any heavy element) are widely 
       scattered at low metallicities.
       The large scatter observed for the abundances of the 
       s-elements at low metallicity (especially for Sr) 
       are the signature of strong inhomogeneities in the interstellar matter 
       which formed the analyzed stars.
       Some metal-poor stars are found with somewhat enhanced `s' or 
      `r'elements, indicating a nucleosynthesis pattern formed just 
        after the onset of the `r' and `s' processes. The r+s rich stars, in
       contrast, show very large enhancements, that have to be 
        explained. 

       %section 4.3. 
       \subsection{The r-process elements in metal-poor stars} \label{r-elem} 

       In a few stars, it happens that the heavy elements are due to the
       r-process only (Burris et al.  2000), very few of these stars are 
       extremely metal-poor : the most extreme examples are CS~22892-052 
       (Sneden et al.  1996, 2003), and CS~31082-001 (Cayrel et al.  2001; 
       Hill et al.  2002).  They have [Eu/Fe]= 1.5 and 1.7 respectively,
       larger than any normal 
       metal-poor stars (Honda, Aoki, Kajino et al.  2004).  
       CS~22892-052 is also carbon-rich. 
       By comparison, the two stars analyzed here, together
       with the one analyzed  by CCO3, have even larger ratios:
       [Eu/Fe] = 1.88, 1.80 and 1.98 dex respectively.  
       Some other C-rich r+s stars observed by 
       AR01, AR02  have also very large [Eu/Fe] ratios : e. g. LP 
       625-44 reaches 2.00 dex. 

       The large [Eu/Fe] ratios observed in the r+s stars 
       could be understood by the addition, to an already r-rich matter, of a 
       large amount of s-rich matter : such a matter includes of course a lot 
       of Ba, but also {\it some} Eu, and a simple calculation shows that, 
       if the exceptional stars strongly enhanced in  r-process 
       elements only,
        receive large amounts of s-process elements, they will reach 
       [Eu/Fe] ratios similar to those of the r+s stars. 
       %(in reverse). 
       AR02 considered that this process of addition could 
       possibly 
       explain the formation of the r+s stars, but the r-only rich very 
       metal-poor 
       stars are so rare that it is implausible to explain a larger 
       population (of r+s stars) through an even rarer one. 

       The imbalance of the two populations would even push towards the 
       reverse scenario (Sect. 4.5). 

       \subsection{The s-process elements in metal-poor stars} 

       The late onset of s-process elements in metal-poor stars is interpreted, 
       according to Burris et al. (2000), by the delay necessary for the 
       ejection 
       of the s elements at the end
        of the evolution of the massive metal-poor AGBs. 
       These AGBs  are presumably formed from 
       interstellar gas enriched in Fe (and possibly r-elements) 
       by the first SN II, and 
       they feed the interstellar gas with s-elements. The
       s-rich non-binary classical metal-poor stars 
       observed would then be third generation stars, since 
       they contain some Fe and s-elements. 
       In some phases (AGB) the low mass sars are able 
       to build themselves s-process elements (and dredge up them to 
       external layers)  : they are then intrinsic 
       s-rich stars. But 
       the analyzed stars are not evolved enough to have reached the 
       AGB phase. They could perhaps have suffered a brief and violent 
       mixing at the 
       core helium flash, and some neutron flux exposure, 
       but the models of He-flash do not produce the expected
       abundance ratios (Weiss et al. 2004).

       In the long period binaries, 
       the enhancement in s-elements has a classical interpretation: mass 
       transfer from a more massive companion in its AGB phase.
       The exceptionally strong enrichment in C, N, and s-elements is
      thus explained by a local event. The stars are extrinsic s-rich stars.
      They may be second generation stars.
      However a variable radial velocity has not yet been found
      for all the few r+s stars. If these stars are found to be non-binaries, 
       they could have been formed from an
       interstellar gas cloud locally enriched by the winds of nearby 
       AGB stars, and this is an extension of the classical extrinsic 
       enrichment (Busso et al. 2001) : it applies to non-binary stars. 
       This explanation of the enrichment may seem somewhat ad hoc, 
       but the scatter of the abundances of the s- and r-elements is large at 
       low metallicities, so large that it implies a clear inhomogeneity of 
       the interstellar matter. 

       The r+s stars not recognized as binaries may also 
       have been initially long period binaries, 
       ceasing to be binaries due to the explosion 
       of the massive companion. 

       %section 4.5. 
       \subsection{Possible origins of the metal-poor r+s enhanced stars} 

       It has been noted in Sect. \ref {r-elem}, that a star with a high 
       ratio r-elements/Fe could,  by accreting some s-rich matter,
        become an r+s star; however, given that metal-poor stars with very high 
       r-elements/Fe ratios are very rare, 
       this is not a plausible scenario for the formation of a number of 
        r+s stars (AR02). 

       The interpretation by Tsujimoto and Shigeyama (2001) 
       of the well observed supernova 1987A predicts that the r-rich ejecta 
       of this rather low mass supernova could nearly fit 
       the abundances of both light metals and r-elements 
       in the r+s stars. It is not impossible that a gas cloud, polluted  
        in r-elements by such low-mass low energy SNe II, producing r-rich 
       stars, to be later polluted by an AGB companion.
       If the strong enrichments in s-elements are explained by local
       events (mass transfer), it is tempting to try to explain 
      (completely or in part) the enhancement in r-elements also by a local 
       enrichment from a companion, in this case producing predominantly r-elements.
  Some scenarios have been proposed:

       {\it (i)} For our r+s stars, a first scenario invoking a triple system (one 
       component polluting with r-elements, and another one polluting with 
       s-elements) may be considered, but is does not seem very likely (CC03),
        moreover such a triple system may be dynamically not stable.\\ 
       ~\\ 
       {\it (ii)} A second scenario (CC03)
        proposes that, after mass loss, 
       the polluting massive AGB companion evolves into a white dwarf, which 
       accretes a substantial amount of mass from the s-enriched giant star, 
       then collapses to a neutron star (Accretion Induced collapse : AIC), 
       producing r-elements ejected towards the surface of the observed star. 
       Some features of this scenario may be considered as drawbacks: 

       a) some of the superficial s-enriched matter leaves the 
        giant to go back to the evolved companion, weakening 
        the previous enrichment. 

       b) for an efficient accretion from the giant to the white 
        dwarf, 
        the orbit has to remain a close one, whereas the long period current 
       orbits 
        are not close, but they may have changed  (maybe by an explosion). 

       c) the production of significant amounts of 
        r-elements by the AIC  process has not yet been completely proven.\\ 
       \\ 
     {\it (iii)}  For a third scenario, Zijlstra (2004) proposes that the primary 
       (evolved as an AGB) transfers s-rich matter to the observed star but 
       does not suffer a large mass loss (owing to its low metallicity).  At 
       the end of the AGB phase, the degenerate core is massive enough for 
       exploding as a supernova of type SN 1.5 (so called by Iben and Renzini 
       1983) now called an AGB supernova.  In this scenario the mass-loss is 
       a crucial issue.  Computations by Herwig for moderately massive AGBs 
       (private communication) would indicate a low mass for the final 
       degenerate core, which would then exclude the collapse as a SN 1.5. 
       Alternatively, massive AGBs building massive cores could be considered. 
       The evolution of the mass of the core (and 
       especially the mass loss rate) are difficult 
       problems, so that  at present, the collapse of a massive AGB 
       (M $\approx$ 10-12 M$_{\odot}$) by electron capture is not excluded 
       (Siess, private communication), perhaps producing r-elements. 
       Such a large mass AGB star could possibly provide the 
       observed enhancement of s-process elements in a first phase, 
       and explode or collapse 
       providing the r-elements (Zijlstra, Siess, private communication), 
       but the modelisation of the evolution of such a large mass 
       metal-poor star is a difficult task (especially the estimation of the 
       mass loss). 

       In all these scenarios, the r-rich mass, which would be transferred to
       the observed star, should not alter the high observed ratio of
       [$\alpha$-elements/Fe].  It could have a very high ratio [r-elements/Fe]
       and could even be devoid of Fe, in agreement with the ideas of 
       Qian \& Wasserburg (2003).  It could also have about the
       same high ratio [$\alpha$-elements/Fe] as the observed star.
       In fact, the observed ratio [$\alpha$-elements/Fe] is particularly
       high (see  Table \ref {T-metalabund}, and Table 3 of CC03), even excluding 
       from the computation the proton capture elements like Na, Al or possibly Mg, 
       which may have been partly synthesized by deep CNO burning in the massive 
       companion.

Finally, Johnson \& Bolte (2004) have recently analyzed a metal-poor star, similar
to the two stars analyzed here. They note that their observations cannot
be completely matched by any combination of s-and r-processes, and
propose an alternative explanation. They try to interpret the abundance
pattern of the heavy elements by theoretical metal-poor s-processes, and
although the general trend from Y to Pb is somewhat better represented
than using the solar process, the abundances of some elements are not
satifactorily explained. Further investigations are obviously needed.

       %section 4.6. 
       \subsection{A possible scenario for the r-only stars ?} 

       A modified scenario could possibly explain the rare and mysterious 
       stars strongly enriched in r-only elements such as CS 22892-052 
       (Sneden et al.  1996, 2003), and CS 31082-001 (Cayrel et al.  2001; 
       Hill et al.  2002).  These stars could possibly be formed in a close 
       orbit version of the scenarios forming the r+s stars, favoring an early
       mass transfer from the massive companion in its RGB phase.
        Admittedly, no 
       variation of radial velocity has yet been observed for one of these 
       two r-only stars, but the explosion which would have provided the r-elements 
       could have also disrupted the binary system.  If the massive companion 
       produces some C (and N) but no s-elements, if its final explosion produces 
       r-elements, this scenario forms, by local enrichment, a r-only rich 
       star, explaining simultaneously the mysterious enrichment in C (and N) 
       sometimes observed in these stars. 
       \\ 
       Finally, the r-only stars could be something like the r+s binaries 
       deprived from s-elements. 

       Probably other scenarios are possible. As already noted, the interpretation 
       by Tsujimoto and Shigeyama (2001) of the supernova 1987A, if 
       confirmed, could explain by local enrichment 
       the abundances of both light metals and r-elements in the 
       two known extremely metal-poor r-only rich stars. 

       %section 4.7. 
       \subsection{Some constraints on the scenarios}

       The first 4 stars of AR02 in Table 5 are very similar to 
       the two stars analyzed here and the one by CC03. Two of them 
       have  not yet been found to be binaries (AR02, AR03)
       If they are confirmed as non-binaries, the scenarios based on
         local inhomogeneities of the star forming gas, or the scenarios
       including an explosion of the massive 
       companion and the disruption of the binary system would be favored
       for such cases. 

       All the scenarios implying a transfer of r-elements from a companion 
       predict the 
       formation of a neutron star (possibly displaced in a wider orbit or 
       ejected). 

       In the future, it would be useful to gather for each star precise 
       radial velocities all along a complete orbit, to measure its precise 
       proper motions (e.  g. by a satellite like Gaia) in order to find the 
       mass of the unseen companion (remnant), the excentricity of the orbit. 
       If the orbital plane is perpendicular to the line of sight, precise 
       proper motions will decide about binarity, and contribute to the
       definition of the orbit. Interferometric  measurements would help. 
       Spectra in the UV could also provide more information about 
       r-elements, and X-ray observations (and even neutrino observations) 
       would be of interest. 

       Finally, it is interesting to note that no
       enhancement of the proton capture elements Na and Al is  observed in 
       field metal poor stars, whereas they are common in
       in globular clusters, and often 
       interpreted as enrichment by the winds of AGB stars, the produced 
       elements being retained in the potential well of the cluster. The two 
       field stars analyzed here show a somewhat similar process. 
       Therefore, enhancement in Na and Al may be an indicative of pollutin
       by an AGB companion, as a general rule.

% Table 6
%\begin{table}
%\caption[1]{[Ba/Eu] and [hs/ls] computed as in Norris et al. (1997a):
%[$<$Ba,Ce,Nd$>$]/[$<$Sr,Y,Zr$>$]} \label{T- hs/ls ratios}
%\begin{flushleft}
%\begin{tabular}{llllcllllll}
%\noalign{\smallskip}
%\hline
%\noalign{\smallskip}
%  star   & [Ba/Eu]  &  [hs/ls] & \\
%\noalign{\smallskip}
%\hline
%\noalign{\smallskip}
%LP~625-44   &  0.82 & 1.39        &      \\
%LP~706-7    &  0.61 & 1.76 - 1.44 &      \\
%CS~22948-27 &  0.06 & 1.26        &      \\
%CS~29497-34 & -0.06 & 0.97        &      \\
%CS~22892-52 & -0.85 (-0.67 to -0.97)& 0.5   &      \\
%Sun        &  0    & 0           &      \\
%HD~122563  & -0.5  & -0.5 to 0   &      \\
%\noalign{\smallskip} \hline \end{tabular}
%\end{flushleft}
%\end{table}

%--- 5 ---
\section{Conclusions}\label{Conclusions}

 A revised analysis was
carried out of the very metal-poor CH/CN-strong stars  CS~22948-27 and 
CS~29497-34, including
 a new derivation of heavy element (Z $>$ 35) abundances.
The abundance patterns are improved, relative to Paper I,
regarding  $\alpha$-elements
and  proton-capture elements,
 especially for the two crucial elements Eu and Pb.

New radial velocity measurements show that both stars are long period
binaries, indicating that mass accretion from the (now faded-out)
massive companion occurred at its AGB phase (not RGB), explaining the
observed high enhancement of the s-elements including Pb.

The abundance patterns of the elements, in particular of the
neutron-capture elements, appear to be similar in CS~22948-27 and
CS~29497-34.  Moreover the two stars are also very similar in terms of
kinematic parameters (Galactic velocities),
as demonstrated in Paper I.
 
The abundance patterns are also very similar to that found in the star
HE~2148-1247 analyzed by CC03, probably also a long period 
variable.

These three r+s stars, together with the first four stars of the list
of AR02, have rather similar abundance patterns of 
heavy elements and, at least from this point of view, may form an
homogeneous group.  They have presumably the same origin, but two of them
 are not yet known to be binaries.  They are not clearly
different from the other progressively less extreme stars analysed by
AR01, AR02.

The abundances of the neutron capture elements in CS~22948-27 and
CS~29497-34 appear to be partly due to s-process and partly due to
r-process, about in the same proportions as the solar ones, similarly
to what is found for HE 2148-1247 by CC03.  Also, the
amount of Eu and Ba in these stars is large, comparable to that found
in the Sun.

The s-elements in these stars, very probably accreted from a massive
companion in its AGB phase, show a large abundance of Pb, in
agreement with the computations of Gallino et al.  (1998) who predict
that low metallicity AGB produce high Pb abundances (although the
prediction is for low mass AGB at contrast to what is expected here).
The dominant isotope should be $^{208}$Pb.  Observations in the r+s stars
and the predictions of recent models show only a partial agreement
(Fig. \ref{F-Goriely}).

The three extreme r+s stars, forming an homogeneous group, show also
some overabundance of the proton capture elements Na, and Al,
presumably formed by deep CNO burning in the massive AGB companion.

The observed r-elements may have been transferred from the massive
companion, if, owing to accretion or small mass loss, its final core
is massive enough for producing r-elements induced by its collapse.

The enrichment by transfer of r-rich matter has been such as to preserve the 
clear enhancement of $\alpha$-elements observed in the three r+s stars.

This local enrichment by r-process elements, proposed for our
r+s stars, could perhaps be tentatively extended to the r-only rich
stars in a modified version : the orbit would be closer, the mass
transfer from the massive companion would be in its RGB phase, the
transfer would bring a limited amount of C (and N), and no s-elements,
  the explosion of
the massive final core would provide the r-elements and would induce
(or not) the disruption of the close binary.

The r-only rich stars show also the classical enhancement of the
$\alpha$-elements, but not the strong enhancements of the proton capture
elements Na and Al found in the r+s stars (and attributed to the AGB
companion).

Further theoretical work is needed, regarding  yields of low mass zero metal SN
II, as well as of metal-poor massive AGB stars, and also about
 the final phase of the core
of metal-poor high mass AGB stars. The observation of a larger sample of
stars is advisable, and careful determinations along the years of the
orbit of each binary could provide the mass of the evolved companion
(remnant), giving useful constraints on the problem of the
r-process site(s) in these binaries.

\begin{acknowledgements}
We are indebted to Falk Herwig, Albert Zijlstra and Lionel Siess for
helpful comments about possible processes of formation of the r+s
stars.
We acknowledge partial financial support from the Observatoire de Paris,
CNPq/CNRS and Fapesp. 
\end{acknowledgements}

%--------------------------------- References -------------

\end{document}